\documentclass[acmtog, screen, nonacm]{acmart}
\AtBeginDocument{%
  }

\setcopyright{none}
\copyrightyear{2018}
\acmYear{2018}
\acmDOI{XXXXXXX.XXXXXXX}

\acmJournal{TOG}
\acmVolume{37}
\acmNumber{4}
\acmArticle{111}
\acmMonth{8}

\let\fmul=\otimes
\newcommand{\fsum}[1]{\textstyle\oint_{#1}}
\let\fdiv=\oslash

\newcommand{\notion}[1]{\emph{#1}}

\usepackage{tikz}
\usetikzlibrary{positioning, arrows.meta, shapes, calc}
\usepackage{svg}

\begin{document}

\title{Towards representation agnostic probabilistic programming}

\author{Ole Fenske}
\authornote{Both authors contributed equally to this research.}
\email{ole.fenske@uni-rostock.de}
\orcid{0009-0001-3055-478X}
\author{Maximilian Popko}
\authornotemark[1]
\email{maximilian.popko@uni-rostock.de}
\author{Sebastian Bader}
\email{sebastian.bader@uni-rostock.de}
\author{Thomas Kirste}
\email{thomas.kirste@uni-rostock.de}
\affiliation{%
  \institution{Institute for Visual and Analytic Computing}
  \city{Rostock}
  \country{Germany}
}

\renewcommand{\shortauthors}{Fenske et al.}

\begin{abstract}
Current probabilistic programming languages and tools tightly couple model representations with specific inference algorithms, preventing experimentation with novel representations or mixed discrete-continuous models. We introduce a factor abstraction with five fundamental operations that serve as a universal interface for manipulating factors regardless of their underlying representation. This enables representation-agnostic probabilistic programming where users can freely mix different representations (e.g. discrete tables, Gaussians distributions, sample-based approaches) within a single unified framework, allowing practical inference in complex hybrid models that current toolkits cannot adequately express.
\end{abstract}



\maketitle

\section{Introduction}
Probabilistic programming languages (PPLs) and toolkits (PPTs) enable practitioners to express complex statistical models and perform Bayesian inference without manually implementing inference algorithms. However, the scope of models that can be defined successfully depends on the mechanisms available for representing distributions. It is interesting to note that several PPLs (e. g. PyMC \cite{salvatier_probabilistic_2016}, Stan \cite{carpenter_stan_2017}, Pyro \cite{bingham_pyro_2019}) tightly couple model representations with specific inference algorithms (e.g., the NUTS sampler for Stan). The PPTs Factorie \cite{factorie}, RXInfer \cite{rxinfer}, Infernet \cite{infernet} provide a wider range of representations, based on the concept of “factors”, but do not offer a standard way to define new representations.

However, efficient computational probabilistic reasoning often demands highly specialized representations. For instance, structured probability spaces \cite{choi_tractable_2015} – distributions over scattered sets – are difficult to represent using “standard” representations based on samples, (sparse) arrays, or parametric mechanisms; instead they rely on probabilistic sentential decision diagrams  \cite{kisa_probabilistic_2014}. For a “universal” probabilistic reasoning library, it would therefore be desirable to be able to add such specialized representations. In addition, one would like the library API to provide a set of operations that allows to formulate all probabilistic computations in a way that is agnostic to the specific representation chosen.

The factor concept \cite{koller_pgm_2009}, which generalizes the concept of distribution functions, is in principle a powerful abstraction tool. Combined with a standard set of factor operations, it is able to decouple model syntax (probabilistic structure) from semantics (computational realization). This allows inference algorithms, like filtering or smoothing, to be formulated as factor expressions independent of the underlying representation. We therefore advocate to provide the “factor” as fundamental abstraction and remove any assumptions about concrete representations of factors in the factor-level API provided by the toolkit.

In this paper, we outline the basic structure of the factor-level API of such a tool and give an intuitive example utilizing this formalism. 

\section{Factors and factor operations}
A factor $f_{XY}$ over random variables $X$ and $Y$ is an abstract mathematical function $f : X \times Y \rightarrow \mathbb{R}$ that assigns a real number to every configuration $(x, y)$. Factors generalize familiar concepts: probability tables for discrete variables, Gaussian distributions  for continuous variables, mixed discrete-continuous (conditional Gaussians) \cite{Lauritzen01121992} and unnormalized potentials are all factors. Crucially, the abstract mathematical definition is intentionally separate from how a factor is represented in a computer.

Five fundamental operations serve as the API for manipulating factors regardless of their representation:
\begin{table}[h]
    \centering
    \caption{Factor operations.}
    \begin{tabular}{ccc}
        \toprule
        Operation & Expression & Usage\\
        \midrule
        Multiplication & $h_{XYZ}$ = $f_{XY} \otimes g_{YZ}$ & Joining \\ 
        Sum-Out & $g_Y = \fsum{X} f_{XY}$ &  Marginalization\\ 
        Reduction & $g_Y = f_{XY}^{(X=x)}$ & Conditioning\\ 
        \midrule
        Division & $h_{XYZ} = f_{XY} \fdiv g_{YZ}$ & Smoothing\\ %
        Addition & $h_X = f_X \oplus g_X$ & Mixture\\
        \bottomrule
    \end{tabular}
    \label{tab:operations}
\end{table}

\noindent
The first three operations are the conventional standard factor operations that form the algebraic core of message passing in factor graphs (e.g. the sum-product algorithm). \notion{Multiplication} combines messages, \notion{Sum-Out} marginalizes variables, and \notion{Reduction} incorporates evidence \cite{sumproduct}. \notion{Division} is a standard operation for probabilistic inference (e.g. smoothing densities\cite{koller_pgm_2009}), but not always provided in PPTs (its result is not normalizable in general). \notion{Addition} – usually only implicitly used inside a PPT – is necessary to compute mixtures of distributions and specifically for realizing the Sum-Out operation when constructing hierarchical factor representations. 

We here concentrate on identifying the core set of operations required for computations that take distributions as input and return distributions as output; this constitutes the probability-theoretic core. There exists a wide range of useful operations that consume distributions but produce other mathematical objects -- such as MAP estimates, mutual information, or entropy -- which would naturally be part of a more comprehensive “standard library”. The design and formalization of such derived operations is outside the scope of this paper. This separation reflects a semantic distinction: core operations are closed under distributions, whereas derived operations compute functionals or optimizers of distributions.


\section{Representation agnosticism}

Factors can use diverse representations: discrete factors might be arrays, hash-tables, or sentential decision diagrams; continuous factors can be parametric (e.g., Gaussian mean/variance) or non-parametric (samples). The separation between syntax and semantics ensures extensibility: new domain-specific representations can be added by implementing only the representation-specific methods for the factor operators, without altering the core framework.

Inference algorithms can then be implemented at the level of the factor operations introduced above (being defined as generic functions at the API level). Variable elimination \cite{koller_pgm_2009} and belief propagation \cite{pearl_probabilistic_1988} for example naturally decompose into sequences of such factor operations and are explicitly designed for probabilistic inference in factor graphs. For example, filtering in state space models involves the factor expressions:
\begin{enumerate}
    \item \textbf{Prediction step}: $p_{X_{t+1}|y_{1:t}} = \fsum{x_t}p_{X_{t+1}|X_t} \otimes p_{X_t|y_{1:t}}$
    \item \textbf{Correction step} (based on new observation $y_{t+1}$): \begin{align*} 
    \phi_{X_{t+1},y_{1:t+1}} &= p_{X_{t+1}|y_{1:t}} \otimes p_{Y_{t+1}|X_{t+1}}^{(Y_{t+1}=y_{t+1})} \\
    p_{y_{t+1}|y_{1:t}} &= \fsum{x_{t+1}} \phi_{X_{t+1},y_{1:t+1}} \\
    p_{X_{t+1}|y_{1:t+1}} &= \phi_{X_{t+1},y_{1:t+1}} \fdiv p_{y_{t+1}|y_{1:t}}\end{align*}
\end{enumerate}
(Uppercase letters in subscripts define a factor's variables, lowercase letters are simply part of the factor name. Note that $p_{y_{t+1}|y_{1:t}}$ is factor with an empty variable list: a simple scalar.)

Here, each operation dispatches to representation-specific implementations. This maintains a
homomorphism between the defined factor algebra and representation algebra
and enables mixing representations within the same model (e.g. discrete factors as finite maps, continuous factors as Gaussians or samples, or hybrid factors as nested representations) and allows at the same time for seamless extension. 

\section{Exploiting representation agnosticism – a toy example}
\begin{figure}[tbh] 
	\includegraphics[scale=0.55]{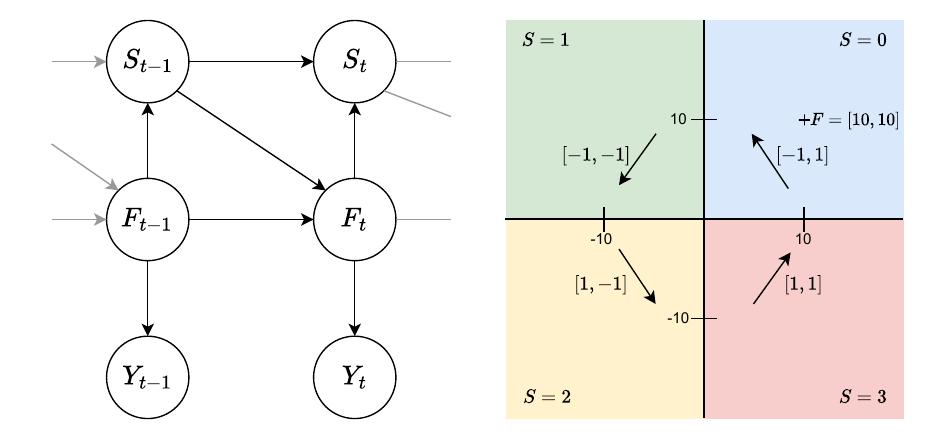}
	\caption{(Left) Graphical representation of dependencies of the model and (Right) an illustration of the quadrants and their linear transition models.}
	\label{fig:fsmodel}
\end{figure}
As a simple example, consider a 2D world partitioned into four quadrants, each quadrant having a primary motion direction, as shown in Fig. \ref{fig:fsmodel}. An object moving in this world will move according to the direction of the quadrant it is in (plus some Gaussian noise). As soon as it crosses the quadrant boundary, it will change motion direction according to the quadrant being entered. This is a simple hybrid dynamic model where a continuous variable $F_t$ (position, range $\mathbb R^2$) interacts with a discrete state $S_t$ (range $\{0,1,2,3\}$, one of four quadrants in the 2D plane).  Figure \ref{fig:fsmodel} illustrates the factorized structure of the quadrant-model, which captures the interaction between the continuous (subsymbolic) state $F_t$ and the discrete (symbolic) state $S_t$ over time. The continuous dynamics $p_{F_t| F_{t-1},S_{t-1}}$
thus describes smooth motion within a region, while the discrete dynamics $p_{S_t| S_{t-1},F_t}$
captures the event of crossing into the next quadrant. The observation factor
$p_{Y_t|F_t}$
links the latent process to the measured data. Together, these factors define the full probabilistic structure of the model. The coupling of continuous evolution and discrete transitions yields a simple hybrid system illustrating how the factorized representation unifies symbolic reasoning with continuous-state estimation: the factor expressions that describe the probabilistic computations are invariant to the kind of random variables involved and the representation chosen.

The main computational challenge arises during the \notion{Multiplication}~($\fmul$) of the conditional Gaussian transition factor
$p_{F_t| F_{t-1},S_{t-1}}$
with the discontinuous link factor
$p_{S_t|S_{t-1},F_t}$. Such a model can not be represented in a simple parametric way. A model developer might be tempted to compare different options for approximate representations – for instance, comparing sample-based representations with parametric approximations using truncated Gaussians and moments matching. 

The \notion{sampling-based representation} approximates these operations via Monte Carlo estimation or particle-based updates. The \notion{parametric representation} expresses factors in closed form (e.g., truncated Gaussian and moments matching) and performs all operations analytically. Both use the same set of generic factor operations, but provide different methods defining their computational realization.


This separation ensures that the probabilistic model can remain fixed, while enabling flexible exploration of different factor representations and inference schemes within a single unified framework.
\begin{figure}[] 
	\includegraphics[scale=0.52]{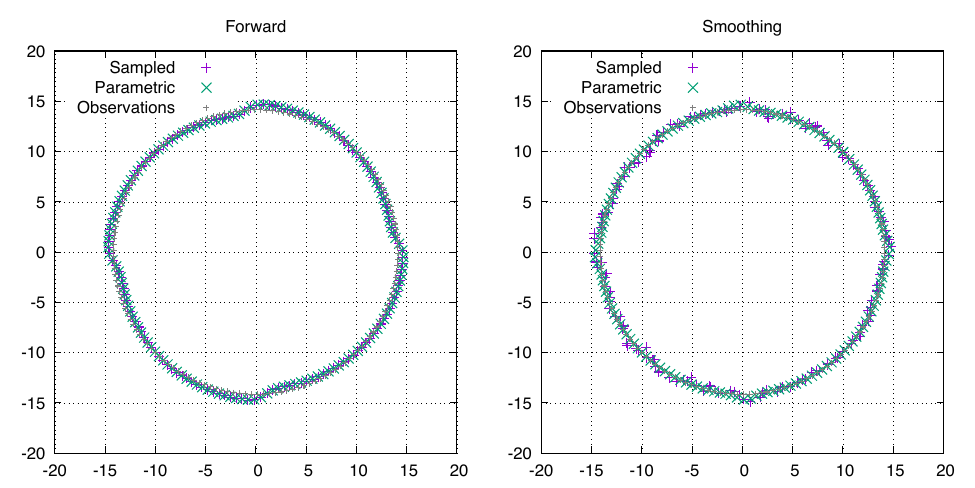}
	\caption{Sampling versus parametric based representation. The mean marginal $F_{1:T}$ state trajectory given circular observations.} 
	\label{fig:result}
    \vspace{-1em}
\end{figure}
Figure \ref{fig:result} shows the filter and smoothing trajectories for both parametric and sampled representations, achieved by simply swapping the underlying factor definition.

\section{Conclusion}

If factor expressions are defined using representation-agnostic generic operations, it becomes possible: (1) to \emph{combine} different representations in the same model, (2) to \emph{exchange} different representations for experimentation, and (3) to add new specialized representations -- without needing to modify a given model.

Our point here is \emph{not} that existing PPLs/PPTs lack object-oriented APIs for probabilistic computations, but rather that \emph{representation agnosticism} at the level of model formulation seemingly has not been an explicit design objective so far -- an objective that we argue should be adopted in future developments.

By treating factors and factor operations as first-class abstractions, we achieve representation-agnostic probabilistic programming where inference algorithms work uniformly across heterogeneous representations. Users can freely experiment with different representation strategies, extend the system with domain-specific representations within a unified framework. This approach enables practical inference in complex hybrid models that current toolkits cannot adequately express or efficiently solve.

As an outlook, it is interesting to consider how computations on distributions relate to sampling-based probabilistic programs. Executing a probabilistic program yields realizations drawn from a (typically joint) distribution. If such realizations are viewed as the dynamic semantics of the program, then the distribution from which they are drawn can be regarded as its static semantics, or semantic type. Under this interpretation, the operations identified in this work form a minimal core for computing the type of a probabilistic program. Recent work such as \cite{faggian_higher_2024} explores the construction of probabilistic program types using typed lambda calculi; our approach connects to this line of research by providing a flexible, distribution-centered foundation for computing and representing such types.

\bibliographystyle{ACM-Reference-Format}
\bibliography{sample-base}



\end{document}